\def\be{\begin{equation}}
\def\ee{\end{equation}}
\def\ba{\begin{array}}
\def\ea{\end{array}}

\documentclass[preprint,showpacs,amsmath]{revtex4}
\usepackage{amsfonts}
\usepackage{graphicx}
\usepackage{amssymb}
\def\qed{\leavevmode\unskip\penalty9999 \hbox{}\nobreak\hfill
     \quad\hbox{\leavevmode  \hbox to.77778em{%
               \hfil\vrule   \vbox to.675em%
               {\hrule width.6em\vfil\hrule}\vrule\hfil}}
     \par\vskip3pt}

\newtheorem{theorem}{Theorem}

\newtheorem{lemma}{Lemma}

\begin{document}
\title{\large\bf Quantum Bell nonlocality cannot be shared under a special kind of bilateral measurements for high-dimensional quantum states}
\author{ Tinggui Zhang$^{1, 2, \dag}$, Qiming Luo$^1$ and Xiaofen Huang$^{1,2}$}
\affiliation{ ${1}$ School of Mathematics and Statistics, Hainan Normal University, Haikou, 571158, China \\
${2}$ Key Laboratory of Data Science and Smart Education, Ministry of Education, Hainan Normal University, Haikou, 571158, China \\
$^{\dag}$ Correspondence to tinggui333@163.com\\}
\date{}
\bigskip

\begin{abstract}
Quantum Bell nonlocality is an important quantum phenomenon.
Recently, the shareability of Bell nonlocality under unilateral
measurements has been widely studied. In this study, we consider the
shareability of quantum Bell nonlocality under bilateral
measurements. Under a specific class of projection operators, we
find that quantum Bell nonlocality cannot be shared for a limited
number of times, as in the case of unilateral measurements. Our
proof is analytical and our measurement strategies can be
generalized to higher dimension cases.

\end{abstract}

  \maketitle

\bigskip
\section{Introduction}
As the source of paradoxes such as the Einstein, Podolsky, Rosen
paradox \cite{qwer} quantum nonlocal correlation was a
controversial phenomenon in quantum mechanics. Nowadays it has become
a key resource in the blooming areas of quantum information and computing \cite{qwet,qwey,qweu,qwei,qweo}. Realizing quantum
violations of the Bell-CHSH inequality \cite{chsh} in various
quantum systems has acquired great interest as evidenced by a wide
range of studies\cite{qwep,qwea,qwes,qwed,qwef,qweg,qweh,qwej,qwek}.
According to quantum physics, measurement outcomes cannot be
predicted with certainty in general \cite{hjki}. Quantum
nonlocality implies that the correlations between the probabilities
of measurement outcomes from two distant systems cannot be
described by classical probability correlation models. Such nonlocal
correlations in multipartite systems have been identified as useful
resources in device-independent quantum information processing
\cite{aaaa}, such as key distribution \cite{cccc,dddd}, randomness
expansion \cite{eeee,ffff,gggg} and randomness amplification
\cite{hhhh}.

Recently, the shareability of quantum Bell nonlocality has been
extensively studied\cite{mnbc,mnbx,ddag,ctdh,mnbv,ztfs,scll}. By
constructing an explicit measurement strategy, the authors in
\cite{mnbv} show that, contrary to previous expectations
\cite{mnbc,mnbx}, there is no limit on the number of independent
Bobs that can have an expected violation of the CHSH inequality with
only one Alice. A class of initial two-qubit states, including all
pure two-qubit entangled states, that are capable of achieving an
unlimited number of CHSH inequality violations has been presented.
This fact has recently been illustrated for higher dimensional
bipartite pure states \cite{ztfs}. Furthermore, in \cite{mnbv}, the
open question of whether quantum nonlocality can be shared under
bilateral measurements was been raised.

In this study, we focus on quantum Bell nonlocality shareability
under bilateral measurements. We consider the following scenario: a
nonlocal correlated bipartite state $\rho_{AB}$ is initially shared
by the first Alice and first Bob. The first Bob performs a randomly
selected measurement, records the measurement outcome, and passes
the post-measurement qubit to the second Bob. Then, the first Alice
performs a randomly selected measurement, records the measurement
outcome, and passes the post-measurement qubit to the second Alice.
The problem of interest is whether the quantum state between the
second Alice and Bob is still nonlocal. In fact, there have been
some numerical and experimental studies on this topic. In
Ref.\cite{scll}, the authors used 17 parameters to verify
numerically that two-qubit quantum states do not share quantum
nonlocality. Moreover, in Ref. \cite{cslb}, they have studied the
sequential generation of Bell nonlocality between independent
observers via recycling the components of entangled systems. They
obtained the stronger one-sided monogamy relations than \cite{scll}.
In Ref.\cite{jzmh}, using entangled photon pairs, the authors
experimentally verified the case of two Alices and two Bobs where
Alice$^{(1)}$ and Bob$^{(1)}$ performed optimal weak measurements
and Alice$^{(2)}$ and Bob$^{(2)}$ performed projective measurements.
To adopt the same measurement strength for Alice$^{(1)}$ and
Bob$^{(1)}$, they observed double EPR steering simultaneously and
showed that double Bell-CHSH inequality violations cannot be
obtained. But for high-dimensional quantum states, the method used
in \cite{scll} is not efficient as too many parameters are involved.
Here, we find that Bell nonlocality cannot be shared under bilateral
measurements for a specific class of projection measurement
operators.


\section{Bipartite state under bilateral measurement}

We considered a measurement scenario where the second Alice
(Alice$^{(2)}$) attempts to share nonlocal correlations of an
entangled pure state with the second Bob (Bob$^{(2)}$)
First, Alice$^{(1)}$ and Bob$^{(1)}$ share an arbitrary
entangled bipartite pure state $|\psi\rangle\in H_A\otimes H_B$,
where $dim(H_A)=s$ and $dim(H_B)=t$ ($s \leq t$). The state has the Schmidt
decomposition form, $|\psi\rangle=\sum_{i=1}^sc_i|i_A\rangle\otimes
|i_B\rangle$, where $c_i\in [0,1]$ and $\sum_i^s c_i^2=1$ and
$\{i_A\}_1^s$ and $\{i_B\}_1^t$ are the orthonormal bases of $H_A$
and $H_B$, respectively. $|\psi\rangle$ is entangled if and only if
at least two $c_i$s are nonzero. Without loss of generality, below
we assume that the $c_i$ are arranged in descending order. The density
matrix corresponding to $|\psi\rangle$ is denoted as
$\rho_{A^{1}B^{1}}=|\psi\rangle\langle\psi|$.

The binary input and output of Alice$^{(k)}$ (Bob$^{(k)}$) are denoted by
$X^{(k)}$ ($Y^{(k)}$) and $A^{(k)}$ ($B^{(k)}$) and $k=1,2$,
respectively. Suppose that Bob$^{(1)}$ performs the measurement according
to $Y^{(1)}=y$ with the outcome $B^{(1)}=b$. Averaged over the
inputs and outputs of Bob$^{(1)}$, the state shared between
Alice$^{(1)}$ and Bob$^{(2)}$ is given by the L\"uders rule
\cite{mnbv}
$$
\rho_{A^{1}B^{2}}=\frac{1}{2}\Sigma_{b,y}(I_s\otimes\sqrt{B_{b|y}^{(1)}})
\rho_{A^{1}B^{1}}(I_s\otimes\sqrt{B_{b|y}^{(1)}}),
$$
where $B^{(1)}_{b|y}$ is the positive operator-valued measure (POVM)
effect corresponding to outcome $b$ of Bob$^{(1)}$'s measurement for
input $y$, and $I_s$ is the $s\times s$ identity matrix. Next
Alice$^{(1)}$ similarly performs the measurement on subsystem A. Then, the state $\rho_{A^{2}B^{2}}$ shared between Alice$^{(2)}$
and Bob$^{(2)}$ is acquired.

To detect the Bell nonlocality of a state $\rho$ we employ the CHSH
inequality \cite{chsh}, $I_{CHSH}=|\langle \mathbb{B}\rangle| \leq
2$, where $\langle \mathbb{B} \rangle = Tr(\mathbb{B} \rho)$,
$\mathbb{B}=A_0\otimes B_0+A_0\otimes B_1+A_1\otimes B_0-A_1\otimes
B_1$, $A_i$, and $B_i$ and $i=0,1$ are Hermitian operators with
eigenvalues of $\in[-1,1]$. If for some binary observables
$A_i^{(k)}$ and $B_i^{(k)}$, $i=0,1$, $I_{CHSH}^{(k)}\equiv
Tr(\mathbb{B}\rho_{A^{k}B^{k}})>2$, then the state
$\rho_{A^{k}B^{k}}$ is nonlocally correlated.

\subsection{Two-qubit pure states}
We first assume that the initial bipartite pure quantum state is a
two-qubit state, $|\psi\rangle\in H_2\otimes H_2$, with Schmidt
decomposition $|\psi\rangle=\sum^2_{i=1}c_i|i_A\rangle|i_B\rangle$.
Namely, $\rho_{A^{1}B^{1}}=|\psi\rangle\langle\psi|$. We employ the
POVMs with measurement operators $\{E,I-E\}$, where $E$ has the form
$E=\frac{1}{2}(I+\gamma\cdot{\sigma_r})$, $r\in R^3$ with $\|r\|=1$,
$\sigma_r=r_1\sigma_1+r_2\sigma_2+r_3\sigma_3$, $\sigma_i$ for
$i=1,2,3$ are the standard Pauli matrices, and $\gamma \in [0,1]$ is
the sharpness of the measurement.

We set the POVM of Alice$^{(1)}$ to \be\label{a00o}
A_{0|0}=\frac{1}{2}(I+(\cos\theta\sigma_1+\sin\theta\sigma_3)), \ee
\be\label{a01o}
A_{0|1}=\frac{1}{2}(I+(\cos\theta\sigma_1-\sin\theta\sigma_3)) \ee
for $\theta\in (0,\frac{\pi}{4}]$. We also let the POVM of
Bob$^{(1)}$ be given by \be\label{b00o}
B^{(1)}_{0|0}=\frac{1}{2}(I+\sigma_1), \ee \be\label{b01o}
B^{(1)}_{0|1}=\frac{1}{2}(I+\gamma_1\sigma_3), \ee where $0\leq
\gamma_1 \leq 1$. Further, we defined the expectation operators $A_x
= A_{0|x} - A_{1|x}$ and $B_y = B_{0|y} - B_{1|y}$ and reached the
following conclusions:

\begin{lemma}
For the quantum state $\rho_{A^{2}B^{2}}$, we have
\begin{eqnarray*}& &Tr[{\rho_{A^{2}B^{2}}}{({\sigma_{1}}{\otimes}{\sigma_{1}})}]\\
&=&{\frac{1+{\sqrt{1-\gamma_1^2}}}{2}}{\cos^2{(\theta)}}Tr[\rho_{A^{1}B^{1}}{({\sigma_{1}}{\otimes}{\sigma_{1}})}]
\end{eqnarray*}
and
\begin{eqnarray*}& &Tr[{\rho_{A^{2}B^{2}}}{({\sigma_{3}}{\otimes}{\sigma_{3}})}]\\
&=&{\frac{1}{2}}{\sin^2{(\theta)}}Tr[\rho_{A^{1}B^{1}}{({\sigma_{3}}{\otimes}{\sigma_{3}})}].
\end{eqnarray*}
\end{lemma}

{\bf{Proof:}} First, after the Bob$^{(1)}$'s measurement we have
$$
\rho_{A^{1}B^{2}}={\frac{2+{\sqrt{1-\gamma_1^2}}}{4}}\rho_{A^{1}B^{1}}+{\frac{1}{4}}{(I\otimes\sigma_{1})}\rho_{A^{1}B^{1}}{(I\otimes\sigma_{1})}+{\frac{1-{\sqrt{1-\gamma_1^2}}}{4}}
{(I\otimes\sigma_{3})}\rho_{A^{1}B^{1}}{(I\otimes\sigma_{3})}.
$$
After Alice$^{(1)}$'s measurement we get
\begin{eqnarray*}
& &\rho_{A^{2}B^{2}}\\
&=&{\frac{1}{2}}{\sum
\limits_{a,x}}{(\sqrt{A_{a|x}}{\otimes}I)}\rho_{A^{1}B^{2}}{(\sqrt{A_{a|x}}{\otimes}I)}\\
&=&{\frac{1}{2}}\rho_{A^{1}B^{2}}+{\frac{1}{2}}
{({\cos{(\theta)}\sigma_{1}}{\otimes}I)}\rho_{A^{1}B^{2}}{({\cos{(\theta)}\sigma_{1}}{\otimes}I)}+{\frac{1}{2}}{({\sin{(\theta)}\sigma_{3}}{\otimes}I)}\rho_{A^1B^{2}}{({\sin{(\theta)}\sigma_{3}}{\otimes}I)}\\
&=&{\frac{2+{\sqrt{1-\gamma_1^2}}}{8}}\rho_{A^{1}B^{1}}+{\frac{1}{8}}{(I\otimes\sigma_{1})}\rho_{A^{1}B^{1}}{(I\otimes\sigma_{1})}+{\frac{1-{\sqrt{1-\gamma_1^2}}}{8}}{(I\otimes\sigma_{3})}\rho_{A^{1}B^{1}}{(I\otimes\sigma_{3})}\\
& &+{\frac{2+{\sqrt{1-\gamma_1^2}}}{8}}{({\cos{(\theta)}\sigma_{1}}{\otimes}I)}\rho_{A^{1}B^{1}}{({\cos{(\theta)}\sigma_{1}}{\otimes}I)}\\
& &+{\frac{1}{8}}{({\cos{(\theta)}\sigma_{1}}{\otimes}I)}{(I\otimes\sigma_{1})}\rho_{A^{1}B^{1}}{(I\otimes\sigma_{1})}{({\cos{(\theta)}\sigma_{1}}{\otimes}I)}\\
& &+{\frac{1-{\sqrt{1-\gamma_1^2}}}{8}}{({\cos{(\theta)}\sigma_{1}}{\otimes}I)}{(I\otimes\sigma_{3})}\rho_{A^{1}B^{1}}{(I\otimes\sigma_{3})}{({\cos{(\theta)}\sigma_{1}}{\otimes}I)}\\
& &+{\frac{2+{\sqrt{1-\gamma_1^2}}}{8}}{({\sin{(\theta)}\sigma_{3}}{\otimes}I)}\rho_{A^{1}B^{1}}{({\sin{(\theta)}\sigma_{3}}{\otimes}I)}\\
& &+{\frac{1}{8}}{({\sin{(\theta)}\sigma_{3}}{\otimes}I)}{(I\otimes\sigma_{1})}\rho_{A^{1}B^{1}}{(I\otimes\sigma_{1})}{({\sin{(\theta)}\sigma_{3}}{\otimes}I)}\\
&
&+{\frac{1-{\sqrt{1-\gamma_1^2}}}{8}}{({\sin{(\theta)}\sigma_{3}}{\otimes}I)}{(I\otimes\sigma_{3})}\rho_{A^{1}B^{1}}{(I\otimes\sigma_{3})}{({\sin{(\theta)}\sigma_{3}}{\otimes}I)}.
\end{eqnarray*}
Then
\begin{eqnarray*}
& &Tr[{\rho_{A^{2}B^{2}}}{({\sigma_{1}}{\otimes}{\sigma_{1}})}]\\
&=&{\frac{2+{\sqrt{1-\gamma_1^2}}}{8}}Tr[\rho_{A^{1}B^{1}}{({\sigma_{1}}{\otimes}{\sigma_{1}})}]+{\frac{1}{8}}Tr[\rho_{A^{1}B^{1}}{({\sigma_{1}}{\otimes}{\sigma_{1}})}]-{\frac{1-{\sqrt{1-\gamma_1^2}}}{8}}
Tr[\rho_{A^{1}B^{1}}{({\sigma_{1}}{\otimes}{\sigma_{1}})}]\\
& &+{\frac{2+{\sqrt{1-\gamma_1^2}}}{8}}{\cos^2{(\theta)}}Tr[\rho_{A^{1}B^{1}}{({\sigma_{1}}{\otimes}{\sigma_{1}})}]+{\frac{1}{8}}{\cos^2{(\theta)}}Tr[\rho_{A^{1}B^{1}}{({\sigma_{1}}{\otimes}{\sigma_{1}})}]\\
& &-{\frac{1-{\sqrt{1-\gamma_1^2}}}{8}}{\cos^2{(\theta)}}Tr[\rho_{A^{1}B^{1}}{({\sigma_{1}}{\otimes}{\sigma_{1}})}]-{\frac{2+{\sqrt{1-\gamma_1^2}}}{8}}{\sin^2{(\theta)}}Tr[\rho_{A^{1}B^{1}}{({\sigma_{1}}{\otimes}{\sigma_{1}})}]\\
& &-{\frac{1}{8}}{\sin^2{(\theta)}}Tr[\rho_{A^{1}B^{1}}{({\sigma_{1}}{\otimes}{\sigma_{1}})}]+{\frac{1-{\sqrt{1-\gamma_1^2}}}{8}}{\sin^2{(\theta)}}Tr[\rho_{A^{1}B^{1}}{({\sigma_{1}}{\otimes}{\sigma_{1}})}]\\
&=&{\frac{1+{\sqrt{1-\gamma_1^2}}}{4}}Tr[\rho_{A^{1}B^{1}}{({\sigma_{1}}{\otimes}{\sigma_{1}})}]+{\frac{1+{\sqrt{1-\gamma_1^2}}}{4}}{\cos^2{(\theta)}}Tr[\rho_{A^{1}B^{1}}{({\sigma_{1}}{\otimes}{\sigma_{1}})}]\\
& &-{\frac{1+{\sqrt{1-\gamma_1^2}}}{4}}{\sin^2{(\theta)}}Tr[\rho_{A^{1}B^{1}}{({\sigma_{1}}{\otimes}{\sigma_{1}})}]\\
&=&{\frac{1+{\sqrt{1-\gamma_1^2}}}{2}}{\cos^2{(\theta)}}Tr[\rho_{A^{1}B^{1}}{({\sigma_{1}}{\otimes}{\sigma_{1}})}].
\end{eqnarray*}

Similarly,
\begin{eqnarray*}
& &Tr[{\rho_{A^{2}B^{2}}}{({\sigma_{3}}{\otimes}{\sigma_{3}})}]\\
&=&{\frac{2+{\sqrt{1-\gamma_1^2}}}{8}}Tr[\rho_{A^{1}B^{1}}{({\sigma_{3}}{\otimes}{\sigma_{3}})}]-{\frac{1}{8}}Tr[\rho_{A^{1}B^{1}}{({\sigma_{3}}{\otimes}{\sigma_{3}})}]+{\frac{1-{\sqrt{1-\gamma_1^2}}}{8}}
Tr[\rho_{A^{1}B^{1}}{({\sigma_{3}}{\otimes}{\sigma_{3}})}]\\
& &-{\frac{2+{\sqrt{1-\gamma_1^2}}}{8}}{\cos^2{(\theta)}}Tr[\rho_{A^{1}B^{1}}{({\sigma_{3}}{\otimes}{\sigma_{3}})}]+{\frac{1}{8}}{\cos^2{(\theta)}}Tr[\rho_{A^{1}B^{1}}{({\sigma_{3}}{\otimes}{\sigma_{3}})}]\\
& &-{\frac{1-{\sqrt{1-\gamma_1^2}}}{8}}{\cos^2{(\theta)}}Tr[\rho_{A^{1}B^{1}}{({\sigma_{3}}{\otimes}{\sigma_{3}})}]+{\frac{2+{\sqrt{1-\gamma_1^2}}}{8}}{\sin^2{(\theta)}}Tr[\rho_{AB^{1}}{({\sigma_{3}}{\otimes}{\sigma_{3}})}]\\
& &-{\frac{1}{8}}{\sin^2{(\theta)}}Tr[\rho_{A^{1}B^{1}}{({\sigma_{3}}{\otimes}{\sigma_{3}})}]+{\frac{1-{\sqrt{1-\gamma_1^2}}}{8}}{\cos^2{(\theta)}}Tr[\rho_{A^{1}B^{1}}{({\sigma_{3}}{\otimes}{\sigma_{3}})}]\\
&=&{\frac{1}{4}}Tr[\rho_{A^{1}B^{1}}{({\sigma_{3}}{\otimes}{\sigma_{3}})}]-{\frac{1}{4}}{\cos^2{(\theta)}}Tr[\rho_{A^{1}B^{1}}{({\sigma_{3}}{\otimes}{\sigma_{3}})}]\\
& &+{\frac{1}{4}}{\sin^2{(\theta)}}Tr[\rho_{A^{1}B^{1}}{({\sigma_{3}}{\otimes}{\sigma_{3}})}]\\
&=&{\frac{1}{2}}{\sin^2{(\theta)}}Tr[\rho_{A^{1}B^{1}}{({\sigma_{3}}{\otimes}{\sigma_{3}})}].
\end{eqnarray*}
\hfill$\Box$

Using the Lemma above, we have the following Theorem:

\begin{theorem}
\smallskip
For any initial entangled bipartite pure quantum state
$|\psi\rangle\in H_2\otimes H_2$,
$|\psi\rangle=\sum^2_{i=1}c_i|i_A\rangle|i_B\rangle$. After the
bilateral measurements, the expected CHSH value of
$\rho_{A^{2}B^{2}}$ is less than or equal to 2, that is,
\begin{eqnarray*}
I_{CHSH}&=&Tr[{\rho_{A^{2}B^{2}}}((A_0+A_1){\otimes}B_0)]\\
& &+Tr[{\rho_{A^{2}B^{2}}}((A_0-A_1){\otimes}B_1) \leq 2.
\end{eqnarray*}
\end{theorem}

{\bf{Proof:}}
\begin{eqnarray*}
I_{CHSH}
&=&Tr[{\rho_{A^{2}B^{2}}}((A_0+A_1){\otimes}B_0)]\\
& &+Tr[{\rho_{A^{2}B^{2}}}((A_0-A_1){\otimes}B_1)]\\
&=&2{\cos{(\theta)}}Tr[{\rho_{A^{2}B^{2}}}{({\sigma_{1}}{\otimes}{\sigma_{1}})}]\\
& &+2{\gamma_1}{\sin{(\theta)}}Tr[{\rho_{A^{2}B^{2}}}{({\sigma_{3}}{\otimes}{\sigma_{3}})}]\\
&=&{\cos^3{(\theta)}}{(1+{\sqrt{1-\gamma_1^2}})}Tr[\rho_{A^{1}B^{1}}{({\sigma_{1}}{\otimes}{\sigma_{1}})}]\\
&
&+{\gamma_1}{\sin^3{(\theta)}}Tr[\rho_{A^{1}B^{1}}{({\sigma_{3}}{\otimes}{\sigma_{3}})}].
\end{eqnarray*}
Since
$Tr[\rho_{A^{1}B^{1}}{({\sigma_{1}}{\otimes}{\sigma_{1}})}]\leq 1$
and $Tr[\rho_{A^{1}B^{1}}{({\sigma_{3}}{\otimes}{\sigma_{3}})}]\leq
1$ we have
\begin{eqnarray*}
I_{CHSH}
&=&{\cos^3{(\theta)}}{(1+{\sqrt{1-\gamma_1^2}})}+{\gamma_1}{\sin^3{(\theta)}}\\
&\leq &2{\cos^3{(\theta)}}+{\sin^3{(\theta)}}.
\end{eqnarray*}

Using
$$f(\theta)=2{\cos^3{(\theta)}}+{\sin^3{(\theta)}},~~~0<\theta\leq{\frac{\pi}{4}},$$
we have
$f'(\theta)=3{\sin{(\theta)}}{\cos{(\theta)}}{[\sin{(\theta)}-2\cos{(\theta)}]}<0$,
as $\sin{(\theta)}<2\cos{(\theta)}$ for
$0<\theta\leq{\frac{\pi}{4}}$. Hence, $f(\theta)$ is a decreasing
function of $\theta$ with $f(\theta)\leq f(0)=2$. Therefore,
$I_{CHSH} \leq 2$. \hfill$\Box$

The above Theorem shows that the second Bob shares no quantum
nonlocality with the second Alice.

\subsection{Generation for Higher dimensional pure states}

For general $d \otimes d$ ($d \geq 3$)  entangled pure state $\rho$
is given by $\rho_{A^{1}B^{1}}=|\varphi\rangle\langle\varphi|$,
where $|\varphi\rangle=\sum_{i=1}^dc_i|ii\rangle$ with
$\sum_{i=1}^dc_i^2=1$. Let
\begin{eqnarray*}
A_{0|0}&=&{\frac{1}{2}}[I_d+\left(
                             \begin{array}{cc}
                               {\cos(\theta)}\sigma_3+{\sin(\theta)}\sigma_1 & 0 \\
                               0 & I_{d-2} \\
                             \end{array}
                           \right)],
\\
A_{0|1}&=&{\frac{1}{2}}[I_d+\left(
                             \begin{array}{cc}
                               {\cos(\theta)}\sigma_3-{\sin(\theta)}\sigma_1 & 0 \\
                               0 & I_{d-2} \\
                             \end{array}
                           \right)],
\\
B_{0|0}&=&{\frac{1}{2}}[I_d+\left(
                             \begin{array}{cc}
                               I_{d-2} & 0 \\
                               0 & \sigma_3 \\
                             \end{array}
                           \right)],
\\
B_{0|1}&=&{\frac{1}{2}}[I_4+\left(
                             \begin{array}{cc}
                               I_{d-2} & 0 \\
                               0 & \gamma_1\sigma_1 \\
                             \end{array}
                           \right)],
\end{eqnarray*}
i.e.
\begin{eqnarray*}
A_0&=&\left(
\begin{array}{cc}
{\cos(\theta)}\sigma_3+{\sin(\theta)}\sigma_1 & 0 \\
 0 & I_{d-2} \\
  \end{array}
   \right)\\
A_1&=&\left(
 \begin{array}{cc}
 {\cos(\theta)}\sigma_3-{\sin(\theta)}\sigma_1 & 0 \\
 0 & I_{d-2} \\
 \end{array}
 \right)\\
B_0&=&\left(
 \begin{array}{cc}
  I_{d-2} & 0 \\
  0 & \sigma_3 \\
  \end{array}
  \right)\\
B_1&=&\left(
\begin{array}{cc}
 I_{d-2} & 0 \\
 0 & \gamma_1\sigma_1 \\
  \end{array}
   \right).
\end{eqnarray*}

Suppose Bob and Alice each perform the measurement above and we
write it as $\rho_{A^{2}B^{2}}$. We can easily obtain the following
Lemma with its proof given in the Appendix:.

\begin{lemma}
For the quantum state $\rho_{A^{2}B^{2}}$, we have
\begin{eqnarray}Tr[\rho_{A^{2}B^{2}}((A_0+A_1){\otimes}B_0)]&=&2\cos^3(\theta)c_1^2-2\cos^3(\theta)c_2^2
-(1+{\sqrt{1-\gamma_1^2}})c_d^2 \nonumber\\
&+&(1+{\sqrt{1-\gamma_1^2}})[c_3^2+c_4^2+\cdots+c_{d-1}^2] \leq
2.\end{eqnarray}
$$Tr[\rho_{A^{2}B^{2}}((A_0-A_1){\otimes}B_1)]=0.$$
\end{lemma}

We can naturally draw the following conclusion.
\begin{theorem}
For any initial entangled bipartite pure quantum state
$\rho_{A^{1}B^{1}}=|\varphi\rangle\langle\varphi|$. After the
bilateral measurements, the expected CHSH value of
$\rho_{A^{2}B^{2}}$ satisfies
\begin{eqnarray*}
& &I_{CHSH}^{(2)}=Tr[\rho_{A^{2}B^{2}}((A_0+A_1){\otimes}B_0)]\\
& &+Tr[\rho_{A^{2}B^{2}}((A_0-A_1){\otimes}B_1)] \leq 2.
\end{eqnarray*}
\end{theorem}
{\bf Remark}: In Lemma 3, there are only the first three terms for
$d = 3$, the last term will appear only when $d \geq 4$.
\section{conclusion and discussion}
In this article, we explored the ability to share the quantum
nonlocality of bipartite quantum states under specific measurements.
It has been shown that in these cases, quantum nonlocality cannot be
shared under bilateral measurements. We have made an attempt in
verifying the shareability of quantum nonlocality for
high-dimensional quantum states. But now our analysis is only true
under the kind of quantum measurements we give. We don't know
whether they are optimal or not. Next, we can discuss the selection
of optimal measurements for bipartite quantum pure states and some
mixed states. For multipartite quantum states, the sharing ability
of nonlocality in unilateral POVM measurement is already very
weak\cite{ztfs,sdsd}, so it should be weaker than bipartite quantum
state in bilateral measurement, and we can continue to study it. In
the latest literature \cite{mhsc}, by characterising two-valued
qubit observables in terms of strength, bias, and directional
parameters, the authors investigated generalising the Horodecki
criterion to nonprojective qubit observables. Therefore, we may
continue to think about a series of problems such as the sharing of
network nonlocality \cite{wxcr} or other quantum resources under
nonprojective measurement. Ref.\cite{scll} discussed that for the
qubit case CHSH nonlocality can be shared by bilateral measurements
when there is a bias on the measurements made by Alice and Bob. It
is an interesting question whether there are similar results for
higher-dimensional quantum states.
\smallskip

\bigskip
{\bf Data Availability Statement:} Our manuscript has no associated
data.

\bigskip
{\bf Acknowledgments}: We thank Shao-Ming Fei, Naihuan Jing and
Yuan-Hong Tao for their helpful discussions. This work was supported
by Hainan Provincial Natural Science Foundation of China under Grant
No.121RC539 and the National Natural Science Foundation of China
under Grant Nos.12126314,12126351,11861031. This project is also
supported by the specific research fund of the Innovation Platform
for Academicians of Hainan Province under Grant No.YSPTZX202215.

\begin{center}
\section*{Appendix}
\end{center}

\noindent{\bf Proof of Lemma 2}

By straightforward calculation we have
$$\rho_{A^{1}B^{2}}={\frac{2+{\sqrt{1-\gamma_1^2}}}{4}}\rho_{A^{1}B^{1}}+{\frac{1}{4}}{(I\otimes\left(
                             \begin{array}{cc}
                               I_{d-2} & 0 \\
                               0 & \sigma_3 \\
                             \end{array}
                           \right))}\rho_{A^{1}B^{1}}{(I\otimes\left(
                             \begin{array}{cc}
                               I_{d-2} & 0 \\
                               0 & \sigma_3 \\
                             \end{array}
                           \right))}$$
                           $$+{\frac{1-{\sqrt{1-\gamma_1^2}}}{4}}{(I\otimes\left(
                             \begin{array}{cc}
                               I_{d-2} & 0 \\
                               0 & \sigma_1 \\
                             \end{array}
                           \right))}\rho_{A^{1}B^{1}}{(I\otimes\left(
                             \begin{array}{cc}
                               I_{d-2} & 0 \\
                               0 & \sigma_1 \\
                             \end{array}
                           \right))}
$$
and
\begin{eqnarray*}
\rho_{A^{2}B^{2}}&=&{\frac{1}{2}}{\sum \limits_{a,x}}{(\sqrt{A_{a|x}}{\otimes}I)}\rho_{A^{1}B^{2}}{(\sqrt{A_{a|x}}{\otimes}I)}\\
&=&{\frac{1}{8}}([I_4+\left(
                             \begin{array}{cc}
                               {\cos(\theta)}\sigma_3+{\sin(\theta)}\sigma_1 & 0 \\
                               0 & I_{d-2} \\
                             \end{array}
                           \right)
]\otimes{I})\rho_{A^{1}B^{2}}([I_4+\left(
                             \begin{array}{cc}
                               {\cos(\theta)}\sigma_3+{\sin(\theta)}\sigma_1 & 0 \\
                               0 & I_{d-2} \\
                             \end{array}
                           \right)
]\otimes{I})\\
& &+{\frac{1}{8}}([I_4-\left(
                             \begin{array}{cc}
                               {\cos(\theta)}\sigma_3+{\sin(\theta)}\sigma_1 & 0 \\
                               0 & I_{d-2} \\
                             \end{array}
                           \right)
]\otimes{I})\rho_{A^{1}B^{2}}([I_4-\left(
                             \begin{array}{cc}
                               {\cos(\theta)}\sigma_3+{\sin(\theta)}\sigma_1 & 0 \\
                               0 & I_{d-2} \\
                             \end{array}
                           \right)
]\otimes{I})\\
& &+{\frac{1}{8}}([I_4+\left(
                             \begin{array}{cc}
                               {\cos(\theta)}\sigma_3-{\sin(\theta)}\sigma_1 & 0 \\
                               0 & I_{d-2} \\
                             \end{array}
                           \right)
]\otimes{I})\rho_{A^{1}B^{2}}([I_4+\left(
                             \begin{array}{cc}
                               {\cos(\theta)}\sigma_3-{\sin(\theta)}\sigma_1 & 0 \\
                               0 & I_{d-2} \\
                             \end{array}
                           \right)
]\otimes{I})\\
& &+{\frac{1}{8}}([I_4-\left(
                             \begin{array}{cc}
                               {\cos(\theta)}\sigma_3-{\sin(\theta)}\sigma_1 & 0 \\
                               0 & I_{d-2} \\
                             \end{array}
                           \right)
]\otimes{I})\rho_{A^{1}B^{2}}([I_4-\left(
                             \begin{array}{cc}
                               {\cos(\theta)}\sigma_3-{\sin(\theta)}\sigma_1 & 0 \\
                               0 & I_{d-2} \\
                             \end{array}
                           \right)
]\otimes{I})\\
&=&{\frac{1}{2}}\rho_{A^{1}B^{2}}+{\frac{1}{4}}(\left(
                             \begin{array}{cc}
                               {\cos(\theta)}\sigma_3+{\sin(\theta)}\sigma_1 & 0 \\
                               0 & I_{d-2} \\
                             \end{array}
                           \right)
\otimes{I})\rho_{A^{1}B^{2}}(\left(
                             \begin{array}{cc}
                               {\cos(\theta)}\sigma_3+{\sin(\theta)}\sigma_1 & 0 \\
                               0 & I_{d-2} \\
                             \end{array}
                           \right)
\otimes{I})\\
& &+{\frac{1}{4}}(\left(
                             \begin{array}{cc}
                               {\cos(\theta)}\sigma_3-{\sin(\theta)}\sigma_1 & 0 \\
                               0 & I_{d-2} \\
                             \end{array}
                           \right)
\otimes{I})\rho_{A^{1}B^{2}}(\left(
                             \begin{array}{cc}
                               {\cos(\theta)}\sigma_3-{\sin(\theta)}\sigma_1 & 0 \\
                               0 & I_{d-2} \\
                             \end{array}
                           \right)
\otimes{I}).
\end{eqnarray*}

Set $P={\cos(\theta)}\sigma_3+{\sin(\theta)}\sigma_1$ and
$Q={\cos(\theta)}\sigma_3-{\sin(\theta)}\sigma_1$.
$\rho_{A^{2}B^{2}}$ can be expressed as
\begin{eqnarray*}
\rho_{A^{2}B^{2}}&=&{\frac{2+{\sqrt{1-\gamma_1^2}}}{8}}\rho_{A^{1}B^{1}}\\
& & +{\frac{1}{8}}{(I\otimes\left(
                             \begin{array}{cc}
                               I_{d-2} & 0 \\
                               0 & \sigma_3 \\
                             \end{array}
                           \right))}\rho_{A^{1}B^{1}}{(I\otimes\left(
                             \begin{array}{cc}
                               I_{d-2} & 0 \\
                               0 & \sigma_3 \\
                             \end{array}
                           \right))}\\
                           & &+{\frac{1-{\sqrt{1-\gamma_1^2}}}{8}}{(I\otimes\left(
                             \begin{array}{cc}
                               I_{d-2} & 0 \\
                               0 & \sigma_1 \\
                             \end{array}
                           \right))}\rho_{A^{1}B^{1}}{(I\otimes\left(
                             \begin{array}{cc}
                               I_{d-2} & 0 \\
                               0 & \sigma_1 \\
                             \end{array}
                           \right))}\\
& &+{\frac{1}{16}}(\left(
                             \begin{array}{cc}
                               P & 0 \\
                               0 & I_{d-2} \\
                             \end{array}
                           \right)
\otimes{\left(
          \begin{array}{cc}
            I_{d-2} & 0 \\
            0 & \sigma_3 \\
          \end{array}
        \right)
})\rho_{A^{1}B^{1}}(\left(
                             \begin{array}{cc}
                               P & 0 \\
                               0 & I_{d-2} \\
                             \end{array}
                           \right)
\otimes{\left(
          \begin{array}{cc}
            I_{d-2} & 0 \\
            0 & \sigma_3 \\
          \end{array}
        \right)
})\\
& &+{\frac{2+{\sqrt{1-\gamma_1^2}}}{16}}(\left(
                             \begin{array}{cc}
                               P & 0 \\
                               0 & I_{d-2} \\
                             \end{array}
                           \right)
\otimes{I})\rho_{A^{1}B^{1}}(\left(
                             \begin{array}{cc}
                               P & 0 \\
                               0 & I_{d-2} \\
                             \end{array}
                           \right)
\otimes{I})\\
& &+{\frac{1-{\sqrt{1-\gamma_1^2}}}{16}}(\left(
                             \begin{array}{cc}
                               P & 0 \\
                               0 & I_{d-2} \\
                             \end{array}
                           \right)
\otimes{{\left(
          \begin{array}{cc}
            I_{d-2} & 0 \\
            0 & \sigma_1 \\
          \end{array}
        \right)
}})\rho_{A^{1}B^{1}}(\left(
                             \begin{array}{cc}
                               P & 0 \\
                               0 & I_{d-2} \\
                             \end{array}
                           \right)
\otimes{{\left(
          \begin{array}{cc}
            I_{d-2} & 0 \\
            0 & \sigma_1 \\
          \end{array}
        \right)
}})\\
& &+{\frac{1}{16}}(\left(
                             \begin{array}{cc}
                               Q & 0 \\
                               0 & I_{d-2} \\
                             \end{array}
                           \right)
\otimes{\left(
          \begin{array}{cc}
            I_{d-2} & 0 \\
            0 & \sigma_3 \\
          \end{array}
        \right)
})\rho_{A^{1}B^{1}}(\left(
                             \begin{array}{cc}
                               Q & 0 \\
                               0 & I_{d-2} \\
                             \end{array}
                           \right)
\otimes{\left(
          \begin{array}{cc}
            I_{d-2} & 0 \\
            0 & \sigma_3 \\
          \end{array}
        \right)
})\\
& &+{\frac{2+{\sqrt{1-\gamma_1^2}}}{16}}(\left(
                             \begin{array}{cc}
                               Q & 0 \\
                               0 & I_{d-2} \\
                             \end{array}
                           \right)
\otimes{I})\rho_{A^{1}B^{1}}(\left(
                             \begin{array}{cc}
                               Q & 0 \\
                               0 & I_{d-2} \\
                             \end{array}
                           \right)
\otimes{I})\\
& &+{\frac{1-{\sqrt{1-\gamma_1^2}}}{16}}(\left(
                             \begin{array}{cc}
                               Q & 0 \\
                               0 & I_{d-2} \\
                             \end{array}
                           \right)
\otimes{{\left(
          \begin{array}{cc}
            I_{d-2} & 0 \\
            0 & \sigma_1 \\
          \end{array}
        \right)
}})\rho_{A^{1}B^{1}}(\left(
                             \begin{array}{cc}
                               Q & 0 \\
                               0 & I_{d-2} \\
                             \end{array}
                           \right)
\otimes{{\left(
          \begin{array}{cc}
            I_{d-2} & 0 \\
            0 & \sigma_1 \\
          \end{array}
        \right)
}}),
\end{eqnarray*}
where $\rho_{A^{1}B^{1}}=|\varphi\rangle\langle\varphi|$.

Therefore,
\begin{eqnarray*}
& &Tr[\rho_{A^{2}B^{2}}((A_0+A_1){\otimes}B_0^2)]\\
&=&2Tr[\rho_{A^{2}B^{2}}(\left(
                          \begin{array}{cc}
                            \cos(\theta)\sigma_3 & 0 \\
                            0 & I_{d-2} \\
                          \end{array}
                        \right)
\otimes\left(
         \begin{array}{cc}
           I_{d-2} & 0 \\
           0 & \sigma_3 \\
         \end{array}
       \right)
)]\\
&=&2Tr[{\frac{2+{\sqrt{1-\gamma_1^2}}}{8}}\rho_{A^{1}B^{1}}(\left(
                          \begin{array}{cc}
                            \cos(\theta)\sigma_3 & 0 \\
                            0 & I_{d-2} \\
                          \end{array}
                        \right)
\otimes\left(
         \begin{array}{cc}
           I_{d-2} & 0 \\
           0 & \sigma_3 \\
         \end{array}
       \right)
)\\
&+&{\frac{1}{8}}\rho_{A^{1}B^{1}}(\left(
                          \begin{array}{cc}
                            \cos(\theta)\sigma_3 & 0 \\
                            0 & I_{d-2} \\
                          \end{array}
                        \right)
\otimes\left(
         \begin{array}{cc}
           I_{d-2} & 0 \\
           0 & \sigma_3 \\
         \end{array}
       \right)
)\\
                           & &+{\frac{1-{\sqrt{1-\gamma_1^2}}}{8}}\rho_{AB^{1}}(\left(
                          \begin{array}{cc}
                            \cos(\theta)\sigma_3 & 0 \\
                            0 & I_{d-2} \\
                          \end{array}
                        \right)
\otimes\left(
         \begin{array}{cc}
           I_{d-2} & 0 \\
           0 & -\sigma_3 \\
         \end{array}
       \right)
)\\
& &+{\frac{1}{16}}\rho_{A^{1}B^{1}}(\left(
                          \begin{array}{cc}
                            P\cos(\theta)\sigma_3P & 0 \\
                            0 & I_{d-2} \\
                          \end{array}
                        \right)
\otimes\left(
         \begin{array}{cc}
           I_{d-2} & 0 \\
           0 & \sigma_3 \\
         \end{array}
       \right)
)\\
& &+{\frac{2+{\sqrt{1-\gamma_1^2}}}{16}}\rho_{A^{1}B^{1}}(\left(
                          \begin{array}{cc}
                            P\cos(\theta)\sigma_3P & 0 \\
                            0 & I_{d-2} \\
                          \end{array}
                        \right)
\otimes\left(
         \begin{array}{cc}
           I_{d-2} & 0 \\
           0 & \sigma_3 \\
         \end{array}
       \right)
)\\
& &+{\frac{1-{\sqrt{1-\gamma_1^2}}}{16}}\rho_{A^{1}B^{1}}(\left(
                          \begin{array}{cc}
                            P\cos(\theta)\sigma_3P & 0 \\
                            0 & I_{d-2} \\
                          \end{array}
                        \right)
\otimes\left(
         \begin{array}{cc}
           I_{d-2} & 0 \\
           0 & -\sigma_3 \\
         \end{array}
       \right)
)\\
& &+{\frac{1}{16}}\rho_{A^{1}B^{1}}(\left(
                          \begin{array}{cc}
                            Q\cos(\theta)\sigma_3Q & 0 \\
                            0 & I_{d-2} \\
                          \end{array}
                        \right)
\otimes\left(
         \begin{array}{cc}
           I_{d-2} & 0 \\
           0 & \sigma_3 \\
         \end{array}
       \right)
)\\
& &+{\frac{2+{\sqrt{1-\gamma_1^2}}}{16}}\rho_{A^{1}B^{1}}(\left(
                          \begin{array}{cc}
                            Q\cos(\theta)\sigma_3Q & 0 \\
                            0 & I_{d-2} \\
                          \end{array}
                        \right)
\otimes\left(
         \begin{array}{cc}
           I_{d-2} & 0 \\
           0 & \sigma_3 \\
         \end{array}
       \right)
)\\
& &+{\frac{1-{\sqrt{1-\gamma_1^2}}}{16}}\rho_{A^{1}B^{1}}(\left(
                          \begin{array}{cc}
                            Q\cos(\theta)\sigma_3Q & 0 \\
                            0 & I_{d-2} \\
                          \end{array}
                        \right)
\otimes\left(
         \begin{array}{cc}
           I_{d-2} & 0 \\
           0 & -\sigma_3 \\
         \end{array}
       \right)
)]\\
&=&Tr[{\frac{3+{\sqrt{1-\gamma_1^2}}}{2}}\rho_{A^{1}B^{1}}(\left(
                          \begin{array}{cc}
                            \cos^3(\theta)\sigma_3 & 0 \\
                            0 & I_{d-2} \\
                          \end{array}
                        \right)
\otimes\left(
         \begin{array}{cc}
           I_{d-2} & 0 \\
           0 & \sigma_3 \\
         \end{array}
       \right)
)\\
& &+{\frac{1-{\sqrt{1-\gamma_1^2}}}{2}}\rho_{A^{1}B^{1}}(\left(
                          \begin{array}{cc}
                            \cos^3(\theta)\sigma_3 & 0 \\
                            0 & I_{d-2} \\
                          \end{array}
                        \right)
\otimes\left(
         \begin{array}{cc}
           I_{d-2} & 0 \\
           0 & -\sigma_3 \\
         \end{array}
       \right)
)]\\
& &=2\cos^3(\theta)c_1^2-2\cos^3(\theta)c_2^2
-(1+{\sqrt{1-\gamma_1^2}})c_d^2
\\
& &+(1+{\sqrt{1-\gamma_1^2}})[c_3^2+c_4^2+\cdots+c_{d-1}^2].
\end{eqnarray*}

Since $2\cos^3(\theta)\leq 2$, $-2\cos^3(\theta)\leq 2$,
$(1+{\sqrt{1-\gamma_1^2}})\leq 2$ and $(1+{\sqrt{1-\gamma_1^2}})\leq
2$, where $ 0<\gamma_1<1, 0<\theta\leq{\frac{\pi}{4}}$, and
$\sum_{i=1}^dc_i^2=1$ we have
$Tr[\rho_{A^{2}B^{2}}((A_0+A_1){\otimes}B_0)]\leq 2.$

Similarly, we have
\begin{eqnarray*}
& &Tr[\rho_{A^{2}B^{2}}((A_0-A_1){\otimes}B_1)]\\
&=&2Tr[\rho_{A^{2}B^{2}}(\left(
                          \begin{array}{cc}
                            \sin(\theta)\sigma_1 & 0 \\
                            0 & 0 \\
                          \end{array}
                        \right)
\otimes\left(
         \begin{array}{cc}
           I_{d-2} & 0 \\
           0 & \gamma_1\sigma_1 \\
         \end{array}
       \right)
)]\\
&=&2Tr[{\frac{2+{\sqrt{1-\gamma_1^2}}}{8}}\rho_{A^{1}B^{1}}(\left(
                          \begin{array}{cc}
                            \sin(\theta)\sigma_1 & 0 \\
                            0 & 0 \\
                          \end{array}
                        \right)
\otimes\left(
         \begin{array}{cc}
           I_{d-2} & 0 \\
           0 & \gamma_1\sigma_1 \\
         \end{array}
       \right)
)\\
& &+{\frac{1}{8}}\rho_{A^{1}B^{1}}(\left(
                          \begin{array}{cc}
                            \sin(\theta)\sigma_1 & 0 \\
                            0 & 0 \\
                          \end{array}
                        \right)
\otimes\left(
         \begin{array}{cc}
           I_{d-2} & 0 \\
           0 & -\gamma_1\sigma_1 \\
         \end{array}
       \right)
)\\
                           & &+{\frac{1-{\sqrt{1-\gamma_1^2}}}{8}}\rho_{A^{1}B^{1}}(\left(
                          \begin{array}{cc}
                            \sin(\theta)\sigma_1 & 0 \\
                            0 & 0 \\
                          \end{array}
                        \right)
\otimes\left(
         \begin{array}{cc}
           I_{d-2} & 0 \\
           0 & \gamma_1\sigma_1 \\
         \end{array}
       \right)
)\\
& &+{\frac{1}{16}}\rho_{A^{1}B^{1}}(\left(
                          \begin{array}{cc}
                            P\sin(\theta)\sigma_1P & 0 \\
                            0 & 0 \\
                          \end{array}
                        \right)
\otimes\left(
         \begin{array}{cc}
           I_{d-2} & 0 \\
           0 & -\gamma_1\sigma_1 \\
         \end{array}
       \right)
)\\
& &+{\frac{2+{\sqrt{1-\gamma_1^2}}}{16}}\rho_{A^{1}B^{1}}(\left(
                          \begin{array}{cc}
                            P\sin(\theta)\sigma_1P & 0 \\
                            0 & 0 \\
                          \end{array}
                        \right)
\otimes\left(
         \begin{array}{cc}
           I_{d-2} & 0 \\
           0 & \gamma_1\sigma_1 \\
         \end{array}
       \right)
)\\
& &+{\frac{1-{\sqrt{1-\gamma_1^2}}}{16}}\rho_{A^{1}B^{1}}(\left(
                          \begin{array}{cc}
                            P\sin(\theta)\sigma_1P & 0 \\
                            0 & 0 \\
                          \end{array}
                        \right)
\otimes\left(
         \begin{array}{cc}
           I_{d-2} & 0 \\
           0 & \gamma_1\sigma_1 \\
         \end{array}
       \right)
)\\
& &+{\frac{1}{16}}\rho_{A^{1}B^{1}}(\left(
                          \begin{array}{cc}
                            Q\sin(\theta)\sigma_1Q & 0 \\
                            0 & 0 \\
                          \end{array}
                        \right)
\otimes\left(
         \begin{array}{cc}
           I_{d-2} & 0 \\
           0 & -\gamma_1\sigma_1 \\
         \end{array}
       \right)
)\\
& &+{\frac{2+{\sqrt{1-\gamma_1^2}}}{16}}\rho_{A^{1}B^{1}}(\left(
                          \begin{array}{cc}
                            Q\sin(\theta)\sigma_1Q & 0 \\
                            0 & 0 \\
                          \end{array}
                        \right)
\otimes\left(
         \begin{array}{cc}
           I_{d-2} & 0 \\
           0 & \gamma_1\sigma_1 \\
         \end{array}
       \right)
)\\
& &+{\frac{1-{\sqrt{1-\gamma_1^2}}}{16}}\rho_{A^{1}B^{1}}(\left(
                          \begin{array}{cc}
                            Q\sin(\theta)\sigma_1Q & 0 \\
                            0 & 0 \\
                          \end{array}
                        \right)
\otimes\left(
         \begin{array}{cc}
           I_{d-2} & 0 \\
           0 & \gamma_1\sigma_1 \\
         \end{array}
       \right)
)]\\
&=&0.
\end{eqnarray*}


\bigskip
{\bf References}
\begin{thebibliography}{99}

\bibitem{qwer} A. Einstein, B. Podolsky, and N. Rosen, Can quantum mechanical description of physical reality be considered complete, Physical Review {\bf 47}, 777 (1935).
\bibitem{qwet} M. A. Nielsen and I. Chuang, Quantum computation and quantum information (2002).
\bibitem{qwey} C. Macchiavello, On the role of entanglement in quantum information, Physica A: Statistical Mechanics and its Applications {\bf338}, 68 (2004),proceedings of the conference
A Nonlinear World: the Real World, 2nd International Conference on
Frontier Science.
\bibitem{qweu} C. H. Bennett and S. J. Wiesner, Communication via one and two-particle operators on Einstein-podolsky-rosen states, Phys. Rev. Lett. {\bf 69}, 2881 (1992).
\bibitem{qwei} A. K. Ekert, Quantum cryptography based on bell's theorem, Phys. Rev. Lett. {\bf 67}, 661 (1991).
\bibitem{qweo} A. Ekert, R. Jozsa, and P. Marcer, Quantum algorithms: Entanglement-enhanced information processing [and discussion], Philosophical Transactions: Mathematical, Physical and
Engineering Sciences {\bf 356}, 1769 (1998).
\bibitem{chsh} J. F. Clauser, M. A. Horne, A. Shimony and R. A. Holt, Proposed experiment to test local hidden-variable
theories,Phys. Rev. Lett. {\bf 23}, 880 (1969).
\bibitem{qwep} S. J. Freedman, and J. F. Clauser, Experimental test of local hidden-variable theories, Phys. Rev. Lett. {\bf 28}, 938 (1972).
\bibitem{qwea} A. Aspect, J. Dalibard, and G. Roger, Experimental test of Bell's inequalities using time-varying analyzers, Phys. Rev. Lett. {\bf 49}, 1804 (1982).
\bibitem{qwes} G. Weihs, T. Jennewein, C. Simon, H. Weinfurter, and A. Zeilinger, Violation of Bell's inequality under strict Einstein locality conditions, Phys. Rev. Lett. {\bf 81}, 5039 (1998).
\bibitem{qwed} M. A. Rowe, D. Kielpinski, V. Meyer, C. A. Sackett, W. M. Itano, C. Monroe, and D. J. Wineland, Experimental violation of a Bell's inequality with efficient detection, Nature (London) {\bf 409}, 791 (2001).
\bibitem{qwef} D. N. Matsukevich, P. Maunz, D. L. Moehring, S. Olmschenk, and C. Monroe, Bell inequality violation with two remote atomic qubits, Phys. Rev. Lett. {\bf 100}, 150404 (2008).
\bibitem{qweg} M. Ansmann, H. Wang, R. C. Bialczak, M. Hofheinz, E. Lucero, M. Neeley, A. D. O'Connell, D. Sank, M. Weides, J. Wenner, A. N. Cleland, and J. M. Martinis,
Violation of Bell's inequality in Josephson phase qubits, Nature
(London) {\bf 461}, 504 (2009).
\bibitem{qweh} J. Hofmann, M. Krug, N. Ortegel, L. Gerard, M. Weber, W. Rosenfeld, and H. Weinfurter, Heralded entanglement between widely separated atoms, Science {\bf 337}, 72 (2012).
\bibitem{qwej} M. Giustina, A. Mech, S. Ramelow, B. Wittmann, J. Kofler, J. Beyer, A. Lita, B. Calkins, T. Gerrits, S. W. Nam, R. Ursin, and A. Zeilinger,
Bell violation using entangled photons without the fair-sampling
assumption, Nature (London) {\bf 497}, 227 (2013).
\bibitem{qwek} B. G. Christensen, K. T. McCusker, J. B. Altepeter, B. Calkins, T. Gerrits, A. E. Lita, A. Miller, L. K. Shalm, Y. Zhang, S. W. Nam, N. Brunner, C. C. W. Lim, N. Gisin, and P. G. Kwiat,
 Detection-loophole-free test of quantum nonlocality, and applications, Phys. Rev. Lett. {\bf 111}, 130406 (2013).
\bibitem{hjki} T. E. Stuart, J. A. Slater, R. Colbeck, R. Renner and W. Tittel, An experimental test of all theories with  predictive power beyond quantum theory, Phys. Rev. Lett. {\bf 109}, 020402 (2012).
\bibitem{aaaa} S. Dutta, A. Mukherjee and M. Banik, Operational characterization of multipartite nonlocal correlations, Phys. Rev. A {\bf 102}, 052218 (2020).
\bibitem{cccc} J. Barrett, L. Hardy, and A. Kent, No signalling and quantum key distribution, Phys. Rev. Lett. {\bf 95}, 010503(2005).
\bibitem{dddd} A. Acin, N, Brunner, N. Gisin, S. Massar, S. Pironio, and V. Scarani, Device-independent security of quantum cryptography against collective attacks, Plys. Rev. Lett. {\bf 98}, 230501 (2007).
\bibitem{eeee} R. Colbeck, Quantum and relativistic protocols for secure multi- party computation, Ph.D. thesis, University of Cambridge, 2007, also available as arXiv:0911.3814.
\bibitem{ffff} R. Colbeck and A. Kent, Private randomness expansion with untrusted devices, J. Phys. A {\bf 44}, 095305 (2011).
\bibitem{gggg} S. Pironio, A. Acin, S. Massar, A. Boyer de la Giroday, D. N.Matsukevich, P. Maunz, S. Olmschenk, D. Hayes, L. Luo,T. A. Manning, and C. Monroe, Random numbers certified by Bell's theorem,
Nature (London) {\bf 464}, 1021 (2010).
\bibitem{hhhh} R. Colbeck and R. Renner, Free randomness can be amplified, Nat. Phys. {\bf 8}, 450 (2012).

\bibitem{mnbc} R. Silva, N. Gisin, Y. Guryanova, and S. Popescu, Multiple observers can share the nonlocality of half of an entangled pair by using optimal weak measurements, Phys. Rev.Lett. {\bf 114}, 250401 (2015).
\bibitem{mnbx} S. Mal, A. Majumdar, and D. Home, Sharing of nonlocalityof a single member of an entangled pair of qubits is not possible by more than two unbiased observers on the otherwing, Mathematics {\bf 4}, 48 (2016).
\bibitem{ddag} D. Das, A. Ghosal, S. Sasmal, S. Mal, and A. S. Majumdar, Facets of bipartite nonlocality sharing by
multiple observers via sequential measurements, Phys. Rev. A {\bf
99}, 022305 (2019).
\bibitem{ctdh}C. Ren, T. Feng, D. Yao, H. Shi, J. Chen, and X.
Zhou, Passive and active nonlocality sharing for a two-qubit system
via weak measurements, Phys. Rev. A {\bf 100}, 052121 (2019).
\bibitem{mnbv} Peter J. Brown and Roger Colbeck, Arbitrarily many independent observers can share the nonlocality of a single maximally entangled qubit
pair, Phys. Rev. Lett. {\bf 125}, 090401 (2020).
\bibitem{ztfs} T. Zhang and S. M. Fei, Sharing quantum nonlocality and genuine nonlocality with independent observables, Phys. Rev. A {\bf 103}, 032216
(2021).
\bibitem{scll}S. Cheng, L. Liu, T. J. Baker, M. J. W. Hall, Limitations on sharing Bell nonlocality between sequential pairs of
observers,  Phys. Rev. A. {\bf 104}, L060201 (2021).


\bibitem{cslb} S. Cheng, L. Liu, T. J. Baker, M. J. W.
Hall, Recycling qubits for the generation of Bell nonlocality
between independent sequential observers, Phys. Rev. A. {\bf 105},
022411 (2022).

\bibitem{jzmh} J. Zhu, M. J. Hu, G. C. Guo, C. F. Li, Y. S. Zhang, Einstein-Podolsky-Rosen steering in two-sided sequential
measurements with one entangled pair,Phys. Rev. A {\bf 105}, 032211
(2022).
\bibitem{sdsd} S. Saha, D. Das, S. Sasmal, D. Sarkar, K. Mukherjee, A. Roy, S. S. Bhattacharya, Sharing of tripartite nonlocality by multiple observers measuring sequentially at one side, Quant. Inf. Process.{\bf 18},
42(2019).
\bibitem{mhsc} M. J W Hall and S. Cheng, Generalising the Horodecki criterion to nonprojective qubit observables, J. Phys. A: Math. Theor. {\bf
55}, 045301 (2022).
\bibitem{wxcr} W. Hou, X. Liu, and C. Ren, Network nonlocality sharing via weak measurements in the extended bilocal scenario, Phys. Rev. A {\bf 105},
042436 (2022).
\end{thebibliography}
\end{document}